\documentclass[10pt]{wlscirep}

\title{Bitcoin Covenants: Three Ways to Control the Future}

\author[1]{Jacob Swambo}
\author[2]{Spencer Hommel}
\author[ ]{Bob McElrath}
\author[ ]{Bryan Bishop}

\affil[1]{King's College London, Department of Informatics}
\affil[2]{Fidelity Center for Applied Technology}

\usepackage{blindtext,graphicx}
\usepackage[absolute]{textpos}
\usepackage{array}
\usepackage{algorithm}
\usepackage{algorithmic}
\usepackage{attrib}
\usepackage{mathtools}

\usepackage{amssymb}
\usepackage{pifont}
\begin{abstract}

A bitcoin covenant is a mechanism to enforce conditions on how the control of coins will be transferred in the future. This work introduces a mechanism to construct a general class of covenants without requiring a change to the consensus rules of bitcoin, in contrast to previous covenant mechanism proposals. An exploration of the broad design space of \textit{deleted-key covenants} (using pre-signed transactions with secure key deletion) is given with security analyses that demonstrate a range of possible security models. While the power derived from a mechanism for covenants is undisputed, the method of implementation is contentious. One purpose of this work is to contribute to that debate by demonstrating what is possible today without introducing new security risks to bitcoin. On the other hand, this work makes a compelling case for what can be gained through a soft-fork upgrade for either a Script-based covenant mechanism \cite{BIP119} or a change to the signature hash (SIGHASH) system \cite{BIP118}. The former has had several approaches proposed previously. The latter, introduced herein, is based on an independent proposal aimed at improving off-chain protocols which can be exapted to enable \textit{recovered-key covenants} through elliptic curve key recovery.

\hspace{0.35cm} The dominant differing factor between a covenant mechanism that uses pre-signed transactions with secure key deletion and those that require soft-fork upgrades is the basis of security for each method. The process of secure key deletion is subject to a trade-off between convenience and security. Secure key deletion requires more procedural overhead for a higher level of security (e.g. by relying on multi-signature pre-signed transactions with more signing keys to add redundancy to the deletion process). Moreover, it is recommended that the deletion occurs quickly, which in a multi-party context necessitates interactivity, a practical downside when compared with other mechanisms. Script-based and recovered-key covenants have a non-interactive enforcement, tighter cryptographic assumptions and bypass the trade-off associated with a key deletion process.

\hspace{0.35cm} Key factors that determine the practicality of covenant mechanisms are discussed, including; the enforcement (activation) process, methods for proving accessibility of funds and whether or not they are bound by a covenant, methods for dynamic fee allocation, the cryptographic assumptions that form the basis of their security, and their feasibility in single-party, hierarchical multi-party and adversarial multi-party contexts. Despite the relative downsides of deleted-key covenants, the class of possible covenants it enables is broad and are most practical for custody protocol design. For example, a deleted-key covenant can be used to create layers of security in addition to key management by enforcing combinations of time-locked rate-limits and pre-defined coin flows through an organization. By comparison, it is shown precisely how soft-fork proposals could improve the practicality and utility of bitcoin covenants for custody protocols and enable some adversarial applications such as payment protocols.

\end{abstract}

\begin{document}


\flushbottom
\maketitle

\thispagestyle{empty}

\section{Introduction}

\subsection{Motivation}

In traditional finance parlance, a financial covenant is a commitment in a debt agreement (or other indenture) that certain activities will or will not be carried out \cite{CovInvestopedia}. It should be understood that what will be demonstrated herein only applies to the enforcement of the \textit{commitments} in the covenant, and not to dispute resolution in the case of a default on the \textit{debt agreement}\footnote{In such a case, the resolution is dependent on a legal framework which may fail in its capacity to enforce its verdict due to the bearer nature of bitcoin which makes it difficult to cease. It remains an open question of whether or not it is possible to rely on the blockchain-based consensus of bitcoin to arbitrate peer-to-peer loan agreements in an automated way.}. Traditional mechanisms for dispute resolution of broken commitments may also apply to debt agreements denominated in bitcoin. However, with bitcoin it's possible to enforce a class of covenants programmatically with a mechanism that is secure under a well-defined set of assumptions. Dispute resolution should not be necessary, as the security of the covenant protocol by definition means that the covenant cannot be breached. 

\vspace{0.1cm}
\noindent \textbf{Definition 1.} A \textit{secure bitcoin covenant} is an unbreakable commitment to a specific set of conditions that apply to the transfer of control of the coins. 
\vspace{0.1cm}

Traditionally, there are two classes of covenants; affirmative (where the conditions of the covenant require some specific action to be performed) and negative (where conditions of the covenant require refraining from specific actions). Typically, these covenants will be related to operational capacity of a company - for example to ensure adequate levels of insurance, or to maintain an interest coverage ratio. A simpler example of a covenant is a restriction of the set of possible recipients for future payments. An ability to enforce arbitrary covenants would increase the viability of Bitcoin as a financial system capable of supporting functions required in modern economies. Even a restricted class of bitcoin covenants could enable new opportunities for novel internet-based  business models. Applications of bitcoin covenants are being explored in the design of custody and payment processing protocols \cite{Swambo2020vaults, BIP119}. Some speculative application domains include inheritance planning, social recovery, layer 2 networks, and cryptocurrency derivatives.

Bitcoin covenants can be achieved in single-party and multi-party settings. Self-imposing a covenant (single-party) can be a useful way to add additional layers of security to the funds in one's custody. For example, a choke-point can be enforced where coins must pass through a sequence of transactions before they can be spent arbitrarily to an address external to the custodian's wallet. 

Multi-party covenant protocols may be useful in joint custody operations where participants don't necessarily trust eachother. Consider a bitcoin hedge fund as an example. Bitcoin could be passed through a pre-defined sequence of covenant transactions that encapsulate rate-limiting time-locks from a fund manager to a group of day traders. Covenant transactions could be prepared that automatically send funds to a safe address by any party if they detect malicious activity by another party trying to access more than their allocated amount for the day. This idea is similar to that presented in \cite{ReVault} but where participants gain stricter enforcement of the transaction sequence by use of a covenant mechanism. It may be too risky to give an employee arbitrary spending power over funds, but desirable to give them some autonomy in the processing of the flow of funds through an organization. This can be achieved with pre-signed transactions but covenant enforcement can act as an additional countermeasure against a coalition of insiders with malicious intent. 

Other applications where enforcing covenants in a multi-party setting could be beneficial include off-chain payment processing protocols that involve collaborative transactions. Protocols which utilise unconfirmed, signed transactions such as non-custodial exchange trading \cite{Arwen}, lightning channels \cite{Poon2016} and discreet log contracts \cite{DLC} may find utility by introducing covenants. Again, a benefit comes from enforcing a transaction sequence that mitigates risks associated with the compromise of private keys. An ability to prove the enforcement of a covenant may increase the incentive for the counter-parties to continue with the protocol since the possible spending paths cannot be altered and they may otherwise lose access to the funds in the protocol.

\subsection{Related Work} 

Previous work has demonstrated how adapting the Script programming language enables implementations of bitcoin covenants. An early proposal, from Möser, Eyal, and Sirer \cite{moeser2016bitcoin} described two applications; a powerful custody tool to handle issues that arise with the irreversibility of bitcoin transactions called a \textit{vault}, and a mechanism to create \textit{poison transactions} which they state is a generally useful mechanism for penalizing double-spending attacks. Proceeding work demonstrated how an alternative extension to the Script language could also enable covenants \cite{Covenants2}, and presented how to construct a vault. That work was implemented on a Bitcoin side-chain called Elements-Alpha. Finally, another variant of a Script extension was proposed \cite{BIP119} which demonstrated a few additional use cases; congestion controlled transactions (which enable payment processors to commit to confirmation of transactions when demand for block space is high to scale through-put), congestion controlled channel factories (to scale confirmations of intensive transactions used to create payment channels in layer 2 networks), and covenants for trustless CoinJoin protocols (for improving the privacy properties of bitcoin). Each of the previously proposed methods for implementing a covenant require some non-trivial modification to Bitcoin-core. Gathering community approval for changes to the consensus rules (even backwards compatible changes) has proven to be a slow process. These proposals demand an analysis of how the required modifications broaden the attack surface for bitcoin to gauge whether this is an acceptable cost for the additional functionality of a new financial tool. These proposals haven't concretely presented the benefits they offer since no comparison has been made with what is already possible currently with bitcoin. 


\subsection{Contributions}

This paper introduces two covenant mechanisms, \textit{deleted-key}
covenants, and \textit{recovered-key} covenants\footnote{While these ideas have been known to the community for a while, this is the most extensive presentation of them, to the best of the authors' knowledge.}. This paper also discusses related work on script-based covenants and compares the three approaches. Deleted-key covenants are the only mechanism that works currently since recovered-key and script-based covenants require a soft-fork upgrade. Thus, this paper contributes to the debate of what value those soft-forks would add by direct comparison with what is currently possible.  

Deleted-key covenants, presented first, are based on the concept of using a signature to commit to a transaction template and the future spending path(s) of bitcoin outputs and then deleting the private key used to compute the signature to remove the ability to create additional spending paths in the future. Since no new signatures can be generated, no new transaction templates can be verified.

By itself, the use of pre-signed transactions as a component in custody and payment protocols is interesting because it enables accessible control over funds without exposing a large attack surface. Imagine keeping signing keys on a highly secured group of distributed hardware devices, while managing a set of pre-signed transactions in an alternative wallet, keeping them available for broadcasting when it is relevant. The theft of pre-signed transactions only exposes a small attack surface; a loss of privacy of sensitive data contained in the transactions (e.g. addresses and amounts), and a potential early broadcast of the pre-signed transaction to disrupt the flow of operations. Compare this with the control an attacker gains from the theft of private keys which enable signatures over arbitrary transaction messages. Coupling the use of pre-signed transactions with a secure mechanism to delete the signing key further limits the exposed attack surface associated with funds. The deleted-key covenants proposal herein was inspired by work from Dr. McElrath who introduced the concept of pay-to-time-locked-signed-transactions (P2TST) and explored their application in bitcoin custody \cite{P2TST}.

The process of key deletion is discussed in detail to demonstrate a wide breadth of available security models. As an example, consider signing keys that are contained on disposable devices, which are destroyed after being used to generate a signature on a valid transaction. Destroying the disposable devices offers an effective method for secure key deletion. Moreover, the supply chain risk that signing devices are compromised before use can be significantly mitigated by using cheap heterogeneous commodity hardware in a multi-signature setting for signing devices from numerous manufacturers and suppliers. In this way the desired level of security can be increased at the cost of more hardware and some procedural overhead.


This work also proposes an alternative covenant mechanism, \textit{recovered-key covenants}. The idea makes use of elliptic curve (EC) public key recovery in which a public key is computed from a message (a specific transaction) and its signature. If the signature is chosen as a pair of nothing-up-my-sleeve (NUMS) quantities then the private key is provably unknown. Compared to deleted-key covenants, this has several advantages that derive from avoiding the need for a secure key deletion process. Unfortunately this mechanism is currently impossible. It would be enabled if a soft-fork upgrade such as BIP-118 is accepted \cite{BIP118}.

A summary of the key comparative results is presented in table \ref{tab:table}. A discussion of these results and all relevant definitions are given in the remainder of the paper, which is structured as follows. Section \ref{sec:Preliminaries} provides some useful definitions of basic concepts for bitcoin covenants. Section \ref{sec:Deleted-Key Covenants} introduces deleted-key covenants with a protocol specification, a security analysis, how to compose covenant transaction trees, a discussion of the class of possible covenants, additional requirements for proof-of-reserves, methods for dynamic fee allocation and the mechanism's interactivity requirements. Section \ref{sec:Recovered-Key Covenants} introduces recovered-key covenants with a presentation of elliptic curve public key recovery, a protocol specification and security analysis, the problem of and solution for the factors that preclude this mechanism, and discussion of some advantages and differences with deleted-key covenants. Section \ref{sec:ScriptCovenants} provides a short overview of script-based covenants and demonstrates how these differ compared to deleted-key and recovered-key covenants. Finally, a conclusion is presented in section \ref{sec:Conclusion}.

\begin{center}
\begin{table}
 \begin{tabular}{||c | c | c | c||} 
 \hline
 Covenant Protocol & \textbf{Deleted-key} & \textbf{Recovered-key} & \textbf{Script-based} \\ [0.5ex] 
 \hline\hline
 Enforcement Conditions & Deposit confirmation, & Deposit confirmation, & Deposit confirmation, \\
 (Activation) & signature commitment (I), & signature commitment (NI) & hash commitment (NI)\\
 & key deletion (I) & & \\
 \hline
 Cryptographic Assumptions & Bitcoin consensus, & Bitcoin consensus, & Bitcoin consensus, \\ 
 (Covenant Enforcement) & ECDLP is hard, & ECDLP is hard & secure hash function \\
 & true RNG, &  &  \\
 & secure key deletion & & \\
 \hline
 Cryptographic Assumptions & Honest depositor or & Honest depositor & Honest depositor \\
 (Delegated Custody) & honest threshold $m$ of enforcers &  &\\
 & for $m$-of-$n$ enforcement policy & & \\
 \hline
 Adversarial Multi-Party & Impractical & Practical & Practical \\  
 \hline
 Proof-of-Reserves &  BIP127 \cite{BIP127} & BIP127 \cite{BIP127} & BIP127 \cite{BIP127}  \\
 & + covenant transaction & + covenant transaction & + covenant transaction \\
 & + commitment signature(s) & + commitment signature & + commitment hash \\
 \hline
 Proof-of-Covenant & (I) Key deletion & (NI) Use NUMS quantities & (NI) Show that  \\
 & during enforcement & for chosen $(r,s)$ & commitment hash derives \\
 & & & from covenant transaction \\ 
\hline
 Unconfirmed Transaction & Disruptive & Safe & Safe \\
 Chain Malleability & & & \\
 \hline
 Dynamic Fee  & Only short chains & Long chains are & Long chains are \\
 Allocation & are practical  & practical & practical \\

 \hline
 Relative Size (bytes) &  104 to 106 & 43 to 106 & 34 \\  
 \hline
 Verification Cost & 1& 1& 0 \\
 (Signature Operations) &&& \\
 \hline
 Soft-Fork Upgrade & Not Required & BIP-118 \cite{BIP118} & BIP-119 \cite{BIP119} \\
 \hline

\end{tabular}
\caption{\label{tab:table} A comparison of covenant mechanisms. NI = Non-interactive. I = Interactive. }
\end{table}
\end{center}

\section{Preliminaries}
\label{sec:Preliminaries}

For an overview of bitcoin transactions and the Script language that enables specifying the locking (or unlocking) programs associated with unspent-transaction-outputs (or inputs), readers are referred to chapters 6 and 7 of Mastering Bitcoin \cite{Antonopoulos:2014:MBU:2695500}. Bitcoin covenants are a commitment to a specific set of conditions that apply to the transfer of control of the coins. Coins are transferred with transactions, and so the mechanism for bitcoin covenants relies on creating a commitment to a transaction that will be broadcast in the future and negating any other possible future transactions. 

\vspace{0.1cm}
\noindent \textbf{Definition 2.} A \textit{covenant transaction} is one that has been or will be irrevocably committed to. It represents a restricted set of conditions that must be met in order consume the unspent-transaction-output (UTxO) from which it spends.   

\vspace{0.1cm}
\noindent \textbf{Definition 3.} A \textit{deposit transaction} is one that generates an UTxO which is accessible by a covenant transaction.

\vspace{0.1cm}
\noindent \textbf{Definition 4.} A covenant mechanism's \textit{enforcement conditions} are the set of conditions required to irrevocably commit a set of coins to be bound by a covenant. 

\vspace{0.1cm}
\noindent \textbf{Definition 5.} An \textit{active} covenant is one for which all of its enforcement conditions have been met.

\vspace{0.1cm}
\noindent \textbf{Definition 6.} \textit{Custodial power} is the capacity to spend coins. Custodial power can be \textit{distributed} according to arbitrary access control structures between several parties.
\vspace{0.1cm}

For each covenant mechanism discussed, custodial power can be delegated to an arbitrary access control structure. This is determined by whomever constructs the deposit transaction. The UTxO of a deposit transaction should specify a target address that commits to the covenant transaction \textit{and} to a set of public keys that determine who is granted custodial power. To distribute custodial power among $k$ participants one can use a $j$-of-$k$ multi-signature locking script where each of $k$ participants provide a public key $Q_{i}$ for $i \in \{1, ..., k\}$. A threshold of $j$ participants must collaborate by generating signatures with their associated private keys $q_{i}$ in order to verify and broadcast the covenant transaction. 

\section{Deleted-Key Covenants}
\label{sec:Deleted-Key Covenants}

\subsection{Overview}

Briefly, a simple implementation of a deleted-key covenant is achieved by specifying a covenant transaction with its desired conditions (for example, forcing a specific destination address), generating a \textit{commitment signature} on this transaction, and deleting the private key used to make the signature. Once the deposit has funded the covenant, the UTxO is entirely controlled by the pre-signed transaction since no new signatures can be produced to verify other transactions spending the same UTxO. The pre-signed transaction can be stored until it is favourable to broadcast it. 

For a deleted-key covenant protocol, the enforcement conditions are thus $(a)$ commit to the covenant transaction with a commitment signature, $(b)$ delete the private key used to generate the commitment signature, and $(c)$ confirm\footnote{A transaction is considered \textit{confirmed} when it has been included in a block that is sufficiently deep to be resistant to chain-tip re-organizations.} the deposit transaction. The covenant only becomes active once each of these conditions are met, though the processes may occur asynchronously. With deleted-key covenants, custodial power is determined by access to the covenant transaction and its commitment signature, and any additional access policy that is specified in the deposit transaction. 

In general it is not possible for one party to prove to another that they have deleted a key. Thus, any party that requires a proof that the covenant is being enforced must participate in generating the commitment signature and delete the private key they used to compute it to convince them-self that the covenant is enforced. Thus, when there are multiple interested parties who require a proof-of-covenant, the deposit output should specify an address that commits to a set of \textit{enforcement public keys} $P_{l}$ for $l \in \{1, ..., n\}$ with an $m$-of-$n$ multi-signature locking script. Then, the covenant is only enforced if $n-m+1$ of the associated private keys $p_{l}$ are deleted. If a party requires unilateral enforcement (without an honesty assumption about the other parties) then an $n$-of-$n$ locking script should be used. Otherwise, the security of the enforcement assumes an honest threshold ($n-m+1$) of parties delete their private key.

\subsection{Protocol Specification}

\begin{figure}
    \centering
    \includegraphics[width=140mm]{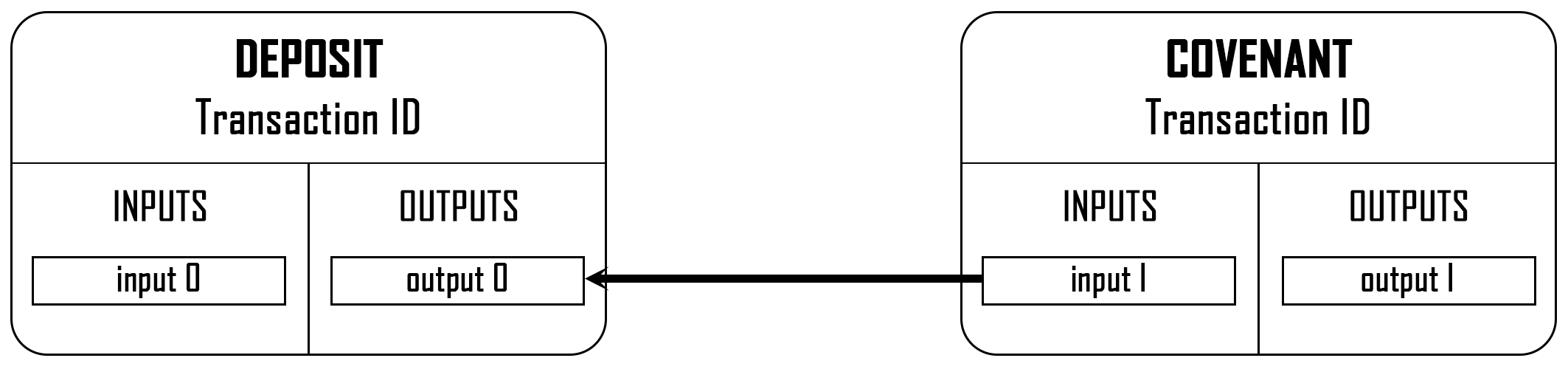}
    \caption{Diagram depicting a simple covenant.  The deposit transaction creates {\tt output 0} which is locked to an address, $A = A(P,Q)$, which is a function of the set of enforcement public keys $P$ and custodial public keys $Q$. {\tt Input 1} of the covenant transaction points to {\tt output 0} from the deposit. A commitment signature is bound to {\tt input 1} of the covenant transaction and commits to the conditions of the covenant, including the transaction version field, the transaction-level lock-time field, and {\tt output 1} which defines the most interesting conditions of the covenant through an arbitrary locking Script.}      
    \label{fig:simple-covenant}

\end{figure}

After completion of this protocol, a set of \textit{custodians} have control over funds which are bound by a covenant (and thus can only spend the funds in accordance with the specific conditions committed to in that  covenant). The protocol assumes the existence of authenticated channels between the \textit{depositor} and all \textit{enforcers}. The construction of the address specified in the deposit assumes prior knowledge of the set of custodial public keys $Q$ but not the set of enforcement public keys $P$. It is assumed that the covenant transaction template is agreed upon prior to enacting its enforcement. A simple example of a deposit and covenant transaction is depicted in figure \ref{fig:simple-covenant}, where both transactions spend from \textit{pay-to-witness-script-hash} (P2WSH) addresses. The address in the deposit transaction's output is a function of the key sets $A = A(P,Q)$.


\begin{enumerate}
    \item Each enforcer generates a key pair $(P_l, p_l)$ and sends $P_l$ to the depositor.
    \item The depositor constructs a Segregated Witness deposit transaction with an UTxO that is locked to the address $A=A(P,Q)$.
    \item The depositor sends details of the deposit transaction (transaction ID, output index, and output amount) to the enforcers.
    \item Each enforcer generates a commitment signature on the covenant transaction, deletes $p_l$, and sends the signature to the depositor with confirmation that the key was deleted.
    \item The depositor verifies each commitment signature. If a quorum of signatures are valid, and the depositor has received notification from each enforcer, the deposit is broadcast. 
    \item The enforcers send their commitment signatures to the set of custodians. 
    \item The depositor sends both the covenant transaction and commitment signatures to the set of custodians.  
\end{enumerate}

Given the use of the Segregated Witness transaction format for the deposit transaction, its transaction ID is independent of the signature and will remain static. This allows for the commitment signatures to be generated, and for the private keys to be deleted \textit{before} committing to the deposit with a signature. This construction is arguably safer than a version of the protocol which doesn't use the Segregated Witness transaction format for the deposit transaction, particularly in a multi-party context. Without Segregated Witness, the deposit transaction would be malleable until it is mined in a block that has reached a sufficient depth to be considered ``confirmed". Thus, the deposit would have to be confirmed before the enforcers commit to the covenant activation by deleting the signing key, otherwise the transaction ID could change and the covenant transaction would be invalid, rendering funds inaccessible. 

\subsection{Security Analysis}

The security of the covenant enforcement depends on how difficult it is to break any of the enforcement conditions. The commitment signature is secure assuming that ECDSA signatures (which are already fundamental to all bitcoin transactions) are secure. The deposit confirmation is secure assuming that bitcoin consensus is robust. General details of the security of bitcoin can be found in other work \cite{Garay2015, Garay2017, Pass2017, Badertscher2017,Badertscher2018}. A more conservative assumption would require that the deposit transaction is deeper in the blockchain and thus less subject to fork-attacks. The key deletion process is the weakest link in the security of covenant enforcement and requires a detailed discussion (see section \ref{subsec:Key Deletion}). For now, note that a wide range of security models are available for the key deletion process.  

There are also risks in the set-up process, before the covenant is active. This protocol assumes the use of a true random number generator (RNG) in the first step when key pairs are generated. There is a high supply-chain risk for any devices used as the source of randomness. Distributing the generation of enforcement key-pairs across a diversity of devices is a reasonable countermeasure for supply-chain risk. It is assumed that the depositor and enforcers communicate over an authenticated, encrypted channel. This means that the enforcement public keys $P_l$ cannot be modified by a man-in-the-middle attack. 

Each enforcement private-key is held by an enforcer for a short period in which it is vulnerable to theft or ex-filtration through both physical and network attacks. Minimizing the lifetime of these private keys is critical. The life-cycle must be at least as long as it takes to enact steps 1-4, where the bottleneck is communication between the set of enforcers and the depositor in step 1. Any enforcer can prolong the life-cycle of the enforcement keys of all others by stalling. During their life-cycle, each private key is vulnerable to network, software, side-channel and hardware attacks. 

A critical countermeasure for mitigating network attacks is to use offline signing devices (such as a hardware wallet) to produce the commitment signature. The hardware wallet typically performs the following functions: key generation, send public key to online device, receive transaction to sign, sign, and send signed transaction. A formal treatment of the security of individual hardware wallets was given in by Arapinis \textit{et al.} in which they demonstrate how to protect against two types of attack; address generation attacks and payment attacks \cite{FormalHardware}. This is achieved by relying on a human to check, through an out-of-band visual communication channel (using the device's screen), that the hardware device is signing the appropriate transaction and that the online device is showing the correct set of addresses. 

Physical security countermeasures should be used to mitigate side-channel and physical inspection attacks. Participants should also maintain network and social privacy to reduce their exposure to external malicious parties. No set of countermeasures will provide perfect protection, but a distributed set of enforcement keys that are generated in heterogeneous (diverse) environments can provide effective security. A strict choice of enforcement policy (e.g. $n$-of-$n$) would mean that an attacker must compromise $n$ distributed offline signing devices before the enforcers delete their signing keys. 

In a multi-party setting, consider the risks posed by malicious parties. If the depositor has malicious intent, they can waste the enforcers' time and computational resources by persuading them to enact the protocol but not broadcasting the deposit. If the enforcers have malicious intent, they can lie about an intention to enforce the covenant and refrain from deleting their enforcement keys. For an arbitrary enforcement policy $m$-of-$n$, the protocol thus assumes that $n-m+1$ enforces will act honestly and delete their keys. 

The motivation for having each enforcer pass their commitment signature to the depositor in step 5 is that the depositor has a natural incentive to forward the commitment signature to (at least some) of the set of custodians once the deposit is broadcast (or else the funds would be lost). Step 6 is a countermeasure against a malicious depositor, who may not transmit the commitment signature to \textit{all} custodians. Thus, custodial power is granted to the full custodian set assuming that either the depositor is honest \textit{or} a threshold of enforcers are honest. For an $m$-of-$n$ enforcement policy, the threshold is $m$.

When the depositor, enforcer and custodian are all the same entity (for example, in a self-custody protocol) the honesty assumptions for enforcement and for delegating custodial power are safe. The honesty assumptions can also be safe in a hierarchical multi-party covenant protocol. For example, consider the example of a hedge fund manager who wishes enforce a covenant and delegate custodial power to a set of day traders while also proving that funds are secure to an insurer. The fund manager's incentive is to act honestly.

However, the honesty assumptions can be impractical in some applications. The worst case is when the depositor and enforcers are distinct from the custodians because there might be no incentive for them to act honestly. If the depositor favours some custodians over others, they may have an incentive to act maliciously. The best case is achieved when the custodians act as enforcers, then they must rely on an honest depositor \textit{or} on an honest threshold $m$ for an enforcement policy $m$-of-$n$ to share their commitment signatures. This demonstrates why adversarial multi-party deleted-key covenants are somewhat impractical. 

The custodians have the responsibility of securing their custodial keys (as is normally the case) \textit{and} the covenant transactions and commitment signatures. The funds in covenant-bound custody are arguably safer than funds controlled by private keys alone, since they are less accessible for arbitrary theft transactions. Another risk for custodians is the malleability of covenant transactions that don't use the Segregated Witness format. This will be discussed further in section \ref{subsec:ClassOfCovenants} as it is relevant for more complex covenant protocols. A malleable covenant transaction could have its transaction ID altered without requiring access to any private keys and thus can make it difficult for the custodian to track their funds. This risk is mitigated by using Segregated Witness covenant transactions.  

When custodial power is distributed, there is a risk that a malicious custodian (or coalition of custodians) broadcast the covenant transaction too early and disrupt the application. A countermeasure for this risk is to use a time-lock in the deposit output such that the covenant transaction can only be broadcast after a specified time. This shifts the security of the application from assuming that the custodian (or coalition) behave honestly and correctly to assuming that the consensus protocol will enforce the time-lock.

It is worth considering how wide-spread use of the protocol could affect the systemic security of bitcoin. Covenants are effectively defined off-chain and a covenant transaction, once broadcast, resembles other common transaction types (P2WPKH, P2WSH, \textit{etc.}) \cite{Antonopoulos:2014:MBU:2695500}. Hence, this covenant mechanism doesn't affect fungibility more than what is already common on the bitcoin network.

Finally, a static fee that has been committed to in the covenant transaction creates operational risks for a covenant-based application. If the covenant transaction specifies a low-priority fee but there is high network demand for block space then there is a chance that the transaction will take a long time to be mined.  Predicting the dynamics of the fee market in advanced is a complex problem, and so methods for dynamically allocating the fee should be used. These methods will be discussed in section \ref{subsec:DKDynamicFee}.


\subsection{Security of Key Deletion}
\label{subsec:Key Deletion}


The security of enforcement of deleted-key covenants depends on the security of private key deletion. The secure erasure of data from a physical medium is a critical security process in many contexts where the disclosure of privileged or confidential information must be avoided. As defined in a survey of the state of knowledge of secure data deletion \cite{SOKSecureDeletion}; `data is securely deleted from a system if an adversary that is given some manner of access to the system is not able to recover the deleted data from the system'. The techniques for securely deleting data differ when using a magnetic hard drive or flash memory. Moreover, the techniques differ when attempting to delete through interfaces at different levels of the system; through a controller at the hardware layer, through the file system at the operating system layer, and through software at the user-interface layer. If the private key's entire life-cycle can be contained in volatile memory (RAM) alone, then its deletion will be dependent on the physical properties of SRAM or DRAM cards. However, if the private key is ever written to disk, its erasure will be dependent on the physical storage medium being used too, typically it will be EEPROM or Flash storage. 

Kissel \textit{et al.} discuss guidlines for media sanitization in \cite{NIST}. They define four types of media sanitization which, in the following order, become more thorough and resistant to information recovery; disposal, clearing, purging, and destroying. Disposal here means simply discarding the storage media and is very susceptible to information recovery. Clearing is a sanitization process for which information cannot be re-gained using data, disk, or file recovery utilities. Overwriting information with random data is an example of clearing. Purging is a sanitization process that resists specialized laboratory attacks. An example of purging from magnetic drives is to use a degausser, which generates a strong magnetic field to remove the information contained in the drive's magnetic domains. Finally, destroying media is the most secure form of media sanitization. This could be done by incinerating, shredding, melting or pulverising the storage media. Factors such as cost must be considered along with effectiveness when deciding which media sanitization processes are suitable. In the context of an organization, the sanitization process should be conducted on a significantly representative sample of storage media to verify the process, ideally employing a non-biased expert entity to conduct the verification.

Valamehr \textit{et al.} \cite{InspectionResistantMemory} discuss physical inspection attacks for determining information contained on storage devices, stating that there are two classes of attacks; \textit{passive} and \textit{intrusive}. Passive attacks will probe the interface of the device to discern timing or electrical differences. Active attacks will breach the boundaries of the device to probe, scan and alter the device internally.  According to Valamehr \textit{et al.} both EEPROM and Flash memory store charge on a floating gate. When their memory cells are overwritten, some residual information is retained as a bias on the substrate. This is the basis for physical inspection attacks with an electron microscope. Similarly for volatile memory, SRAM memory cells can retain some level of information and are subject to similar analyses. Their work explores how to construct device architectures that resist information leakage even when subjected to physical inspection attacks. 

Other work, which also relies on the imperfect nature of recovering data, attempt to incorporate an additional layer of security by using key hiding techniques based on leakage-resistant cryptography. Examples include  exposure-resilient functions and all-or-nothing transforms \cite{ERFAON} and distributed public key schemes secure against continual leakage \cite{DPKSSecureAgainstLeakage}. This exemplifies only a small portion of the work being done to improve physical inspection security of devices, but also demonstrates that the current state-of-the-art fails to handle the totality of the complexity of secure key deletion. These and similar methods may prove useful in future if the theory and implementations advance to support better performing systems and hardware wallet manufacturers can integrate them into their products.

Other work focuses on \textit{provable} deletion for embedded devices \cite{ProofsOfSecureErasure, EPoSE, DeletingSecretData}. The `black box' nature of secure deletion functions, which are often carried out in Trusted Platform Modules (the standard, for example, of key management in the financial industry \cite{Anderson:2008:SEG:1373319}) is unsatisfactory as it requires complete trust in the correct implementation of software inside the module. Provable deletion on the other hand is a scheme to enable users to verify the deletion operations. The provable deletion schemes of \cite{ProofsOfSecureErasure, EPoSE} work by requiring the embedded device to compute a function on a data-set that fills its storage capacity, and return the result as a proof that it stored that data-set which necessitated overwriting all other data previously contained on the device. These techniques may prove very useful for a hardware wallet manufacturer who wishes to enable deleted-key covenant functionality. 

Trusted erasable memory (using secure deletion tools and physical device security measures) and disposable erasable memory (renders physical inspection useless through destruction) differ in their security guarantees and cost effectiveness. Even while using state-of-the-art in trusted erasable memory, one can never be totally certain that no important confidential information has been retained. However, the fact that these devices can be re-used indefinitely for as many secure deletions as is desired makes them more conducive to use in covenant-based applications. Disposable memory offers maximally secure key deletion, but requiring a signing device to be destroyed for each activation of a covenant is impractical. An example of how to minimize wasted resources here would be to create batches of covenants on a simple, cheap and `bare bones' hardware wallet such as a smart card, and to destroy it, activating multiple covenants at a smaller cost than one per card.  

No single method will ever enable perfect key deletion. However, the key deletion process can be distributed across numerous heterogeneous devices which rely on different methods. Each new device and deletion method acts as an additional countermeasure that increases the cost for an attacker. Thus, the security of key deletion can be enhanced by the set of enforcers and the effort they put in to mitigate information recovery attacks. For an enforcement policy of $m$-of-$n$, the key deletion process can resist up to $n-m+1$ private keys being recovered. The trade-off is that more security costs more money and time.

Thus it has been shown that there is a range of possible designs for secure key deletion, each with a different security model and set of assumptions. If security is paramount, then disposable (destructible) memory should be used. However, if probabilistic security is sufficient, a distributed set-up of trusted erasable memory devices may be used.

\subsection{Composability of Covenant Deposits and Commitments}
\label{subsec:Composability}

The design space for covenant-based protocols can be extended from the simple concept depicted in figure \ref{fig:simple-covenant} by increasing the number of inputs and outputs in a covenant transaction and by creating multiple dependent covenant transactions. Additional covenant commitments can be \textit{joint} in the sense that each commitment must be satisfied, or can be \textit{disjoint} such that only a subset of commitments must be satisfied.

\begin{figure}
    \centering
    \includegraphics[width=120mm]{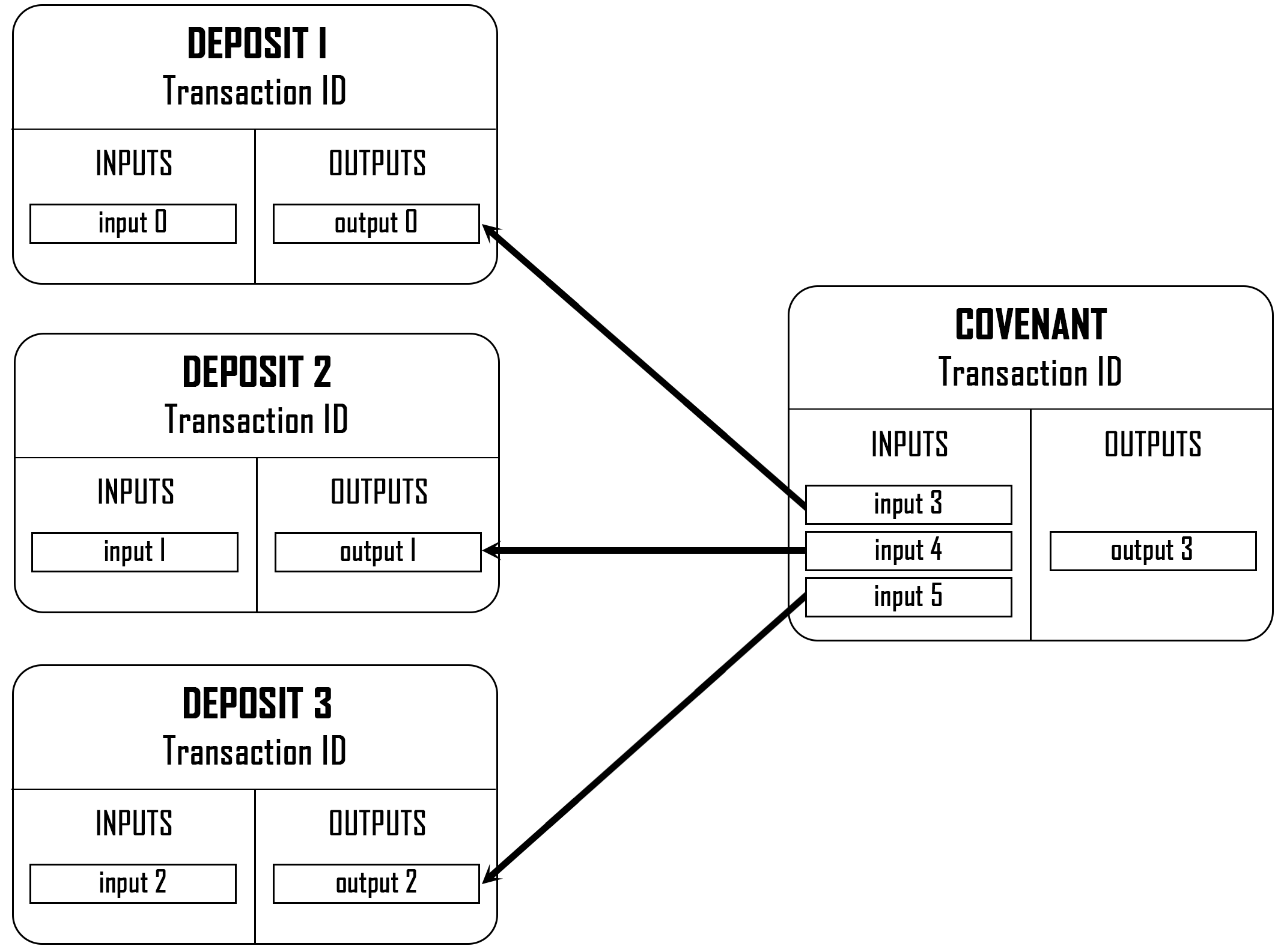}
    \caption{A covenant transaction with three inputs that spend from outputs in different deposit transactions. Each output commits to an enforcement public key and so each input requires at least one commitment signature. All deposits must be confirmed for the signed covenant transaction to become active.}      
    \label{fig:multi-deposit-covenant}
\end{figure}

Figure \ref{fig:multi-deposit-covenant} depicts a covenant transaction with multiple deposit transactions, each of which participate in enforcing the covenant by committing to the public key and its commitment signature(s) associated with the each input. While this is possible, there is a risk that if any of the deposits fail to be confirmed on the blockchain, the funds could be lost. For a self-imposed covenant protocol, relying on Segregated Witness (non-malleable) deposits and the availability and liveness guarantees of the bitcoin network should suffice. In a multi-party setting, it is safer to construct deposits with an additional time-locked refund spending path. The enforcement conditions for deleted-key covenants with multiple deposits require \textit{all} deposits to be confirmed, so this prevents any party from indefinitely locking funds of another by aborting with their deposit.

\begin{figure}
    \centering
    \includegraphics[width=150mm]{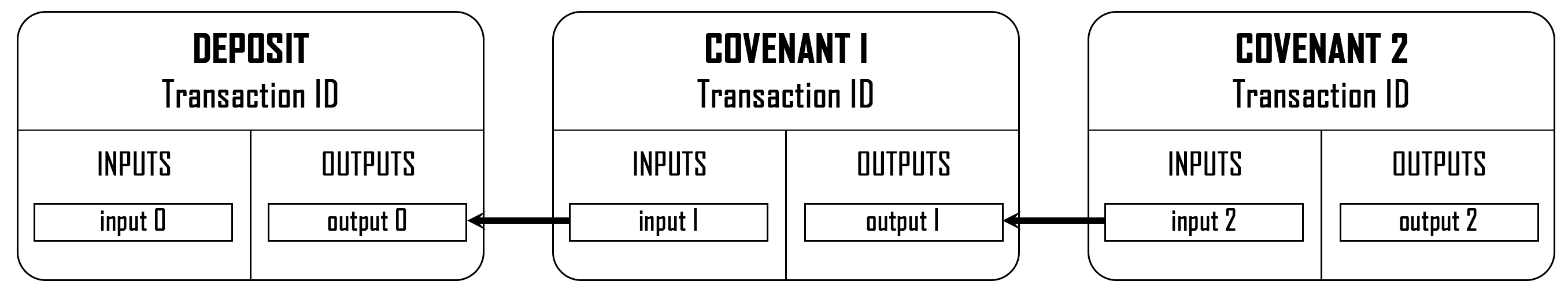}
    \caption{A chain of covenant transactions whose conditions are \textit{joint} in that each must be satisfied to spend the funds which they control. All covenant transactions with dependent transactions must be non-malleable to protect their dependent transactions from being invalidated.}      
    \label{fig:chain-joint-covenant}
\end{figure}

Figure \ref{fig:chain-joint-covenant} depicts joint covenant commitments by constructing a chain of dependent covenant transactions. It is critical to understand \textit{malleability} of transactions when considering off-chain protocols. In BIP-62, a malleable transaction is defined as one which is valid and can be modified in-flight, without invalidating it, but without access to the relevant private keys \cite{BIP62}. Several sources of malleability were elaborated in BIP-62 and most of these sources are handled with the Segregated Witness soft-fork \cite{BIP141, BIP143, BIP144}. Non-malleability is a requirement when constructing an unconfirmed dependent chain of covenant transactions. If the initial transaction's ID is altered, then the references made in each input of subsequent transactions become invalid. For this reason, the Segregated Witness format should be used for covenant-based protocols that use chains of covenant transactions. Here, Covenant 1 must be broadcast before Covenant 2 would be considered valid. Therefore, the conditions encoded by both transactions must be met in order to satisfy the full covenant and regain arbitrary spending capability with the funds. Another way to achieve joint commitments would be to stack multiple spending conditions into the output script of a single covenant transaction.

\begin{figure}
    \centering
    \includegraphics[width=120mm]{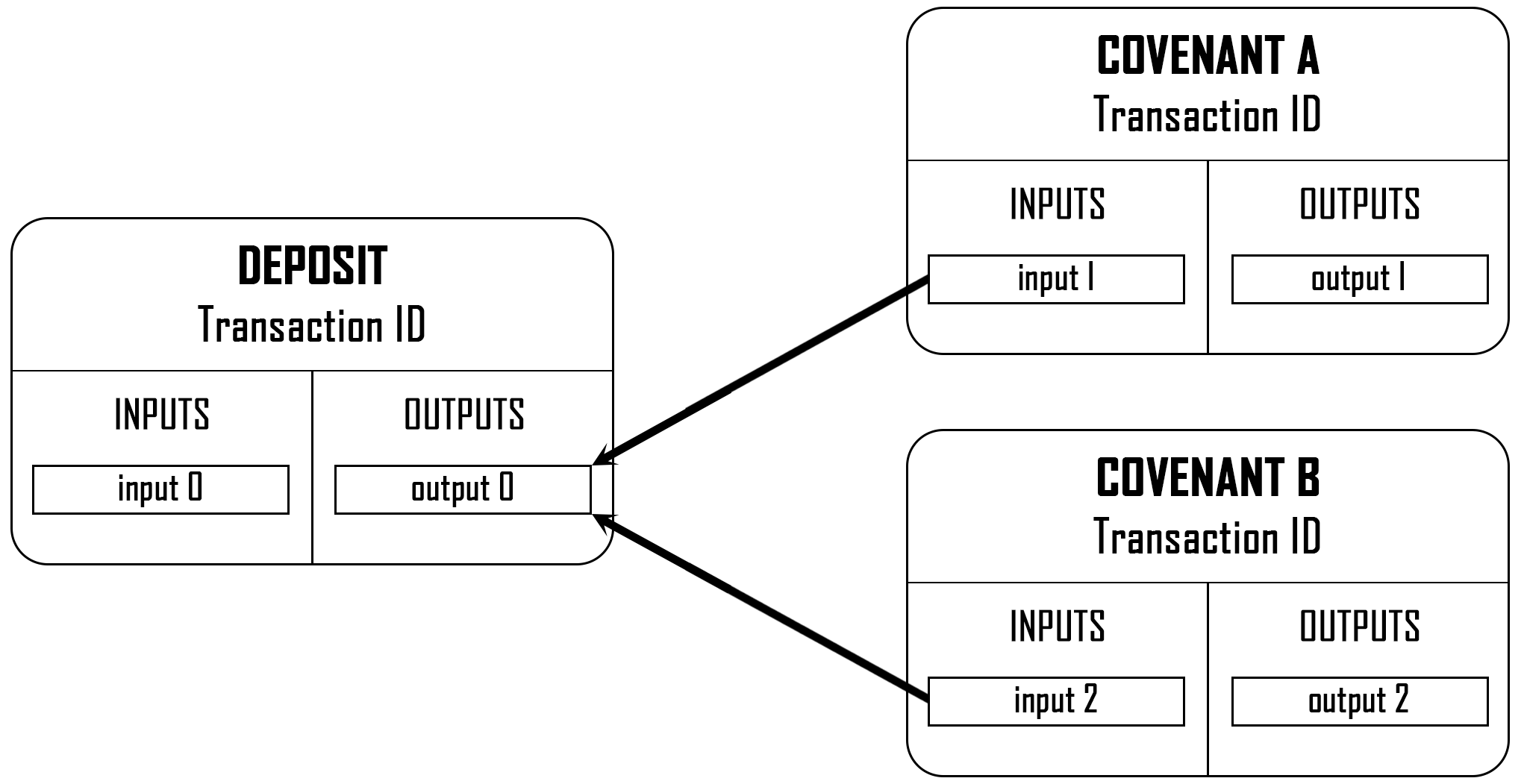}
    \caption{Two covenant transactions which spend from the same deposit output are considered to have \textit{disjoint} commitments in that either covenant A or B must be satisfied while the other is invalidated.}      
    \label{fig:disjoint-covenant}
\end{figure}

Figure \ref{fig:disjoint-covenant} depicts disjoint covenant commitments.There is optionality with which set of conditions to adhere to, either those encoded by Covenant A or by Covenant B. The confirmation of either transaction invalidates the other. Another way to achieve a disjoint commitment would be to use conditional logic in the output locking script of a single covenant transaction. 

Each of these concepts (multiple deposits, joint and disjoint covenant commitments) can be combined arbitrarily to produce multiple dependent chains of covenant transactions in a tree-like structure. Moreover, covenant transactions can be constructed with arbitrary numbers of inputs and outputs provided that transaction size limits are adhered to. One could view these arbitrary combinations as a dynamic and programmed flow of access control for sets of coins.

While an infinitely recursive covenant transaction is not possible with this method, a chain of covenant transactions of arbitrary length is possible. The chain must be predefined since each covenant transaction depends on a previous transaction. Creating the unsigned transactions requires pre-computing the enforcement public keys and knowing the custodial public keys. The enforcement conditions for a recursive covenant require each of the covenant transactions to be signed and each of the enforcement keys to be deleted. Each enforcer must be involved from the beginning and the amounts for each transaction and the specific conditions of each covenant commitment must be known in advance.

While arbitrary trees of covenant transactions would be considered valid according to the consensus protocol of bitcoin and could be directly included in a block by a miner, there are currently practical limitations imposed by policies in the peer-to-peer network that must be considered. For example, bundles of transactions that use `child-pays-for-parent' (CPFP) are not fully supported and will be excluded from peers' memory-pool. Thus, relying on the peer-to-peer network for distributing and executing covenant transaction programs may not be feasible without updating relay policies in the network layer. 

\subsection{Class of Possible Covenants}

\label{subsec:ClassOfCovenants}

It should be clear that the class of possible covenants is predominantly restricted by the parameters of a transaction template. This means that traditional covenants such as those based on the operational capacity of a business could only be implemented by use of an \textit{oracle} who would provide \textit{external} validation of the covenant (the logic of the conditions committed to in the covenant are not understood by the consensus network of bitcoin). Without oracles, deleted-key covenants can have commitments that require only \textit{internal} validation, since each bitcoin node has the logic to independently validate transactions (and transaction chains). 

Constructing a condition that relies on external validation requires participation from an oracle but this introduces some trust assumptions that weaken the security model of the covenant. Script does not enable explicit requests to external servers, nor does it enable importing mutable state variables. A mechanism for enabling an oracle is to give them a necessary signing key in the custodial access control structure, and design them to only provide the signature to verify the transaction once some condition in the external world has been met. Some issues that arise with the use of oracles include; they can be hacked, bribed, or be poorly designed and return incorrect outputs, or they can go offline and a required private key can vanish leading to a loss of funds. Well designed oracles may rely on a consortium of independent signatories to introduce Byzantine fault-tolerance. Moreover, signatories may be instantiated on secure hardware for trusted computing. A more complicated and speculative mechanism for enabling an oracle could rely on multi-party computations (MPCs) to enable arbitrary logic for the access control of funds. The function that is evaluated in the MPC could be considered an oracle which outputs a value which determines whether or not the funds are accessible. Note that the security of this approach would depend on the correct construction of the MPC in addition to the security of consensus in bitcoin and of the key deletion process. 


Covenants that require only internal validation are more practical. Bitcoin transaction signatures are flexible in what parts of the transaction they commit to. By default a signature commits to the transaction version, a proper subset of the inputs and outputs, and the transaction locktime. By using a signature hash (SIGHASH) flag with the signature of each input, signatories can specify which parts of the transaction they endorse. A signature can endorse; all outputs (SIGHASH = {\tt ALL}), a single output (SIGHASH = {\tt SINGLE}), or no outputs (SIGHASH = {\tt NONE}). Additionally, a signature can concatenate these SIGHASH types with {\tt ANYONECANPAY}, which enables the endorsement of only the single input that it comes with, while allowing for other inputs to be added to the transaction. Users typically sign their transactions using {\tt ALL} since this commits to the entire set of inputs and outputs and reduces risks of their signed inputs being `re-packaged' into different transactions. This type of attack is referred to as \textit{signature replay}. The other SIGHASH types are useful in multi-party contexts when a transaction is being passed between multiple participants who may add inputs, outputs, and signatures along the way. Another common use of alternative SIGHASH types is to use {\tt ANYONECANPAY|ALL} to enable adding additional inputs which can pay an increased fee to miners when the network is congested. 

A minimally constrained deleted-key covenant would commit to the version, locktime, and a single input in the transaction (using {\tt ANYONECANPAY|NONE}). The valid input could be placed into different transactions so long as they conform to the format that was signed (i.e. the same version, locktime, and input). A minimally constrained covenant could be a useful feature, however note that any signature that uses a flag other than {\tt ALL} is subject to signature replay which enables malleability attacks that could disrupt covenant-based (and other off-chain) protocols. A maximally constrained covenant would use the {\tt ALL} flag to commit to the version, all inputs, all outputs, and the lock-time. Most of the power of this type of covenant comes from the locking script associated with each output. One can specify that a given set of inputs can only be spent to a set of output addresses whose amounts are well defined and whose unlocking scripts are any set of conditions specifiable by bitcoin Script. Bitcoin Script programs enable conditional payments that must adhere to absolute and relative time-locks \cite{BIP65}, arbitrary access control structures based on public key cryptography\footnote{Using script-based multi-signatures or off-chain techniques such as multi-party threshold signatures schemes \cite{Canetti2020, Gennaro2020,Damgard2020,Gagol2020}.}, secret knowledge, and combinations thereof \cite{BIP199}. Any non-standard Script must be encapsulated in a pay-to-script-hash (P2SH) or pay-to-witness-script-hash (P2WSH) output \cite{BIP16} since by default bitcoin nodes do not relay non-standard transactions. These P2SH and P2WSH outputs pay to addresses derived from a hash of the locking script, called the \textit{redeem script}, which must be provided as input during the signing process. If the redeem script is lost it may be difficult to re-create and funds could get permanently locked.    

The relative storage required for the enforcement of a deleted-key covenant compared with a regular transaction is 104 to 106 bytes. This includes the DER encoded commitment signature (70 to 72 bytes) \cite{BIP66}, the {\tt OP\_CHECKSIG} operation (1 byte) and the compressed enforcement public key (33 bytes). Note that the custodial access policy is not included here.


\subsection{Proof-of-Reserves}
\label{sec:ProofOfReserves}

It is sometimes necessary to demonstrate that one has access to funds using a proof-of-reserves protocol. For example, a custodian that needs to prove solvency to an insurer. Funds that are bound by deleted-key covenants are controlled by a partially-signed transaction (with the commitment signature) and an arbitrary custodial access policy. A custodian can prove that they have access to the private keys associated with the access policy by signing an invalid transaction according to the proof-of-reserves protocol \cite{BIP127}. They should also share the covenant transaction with its commitment signatures and demonstrate that the address derives from the set of enforcement keys and custodial keys.

\subsection{Dynamic Fee Allocation}
\label{subsec:DKDynamicFee}

Since the fee market is dynamic and unpredictable, and funds may be controlled by covenants for long periods of time, there will likely be a mismatch between the fee specified in the covenant transaction and the optimal fee based on the network usage when the time to broadcast it comes. There are two mechanisms in bitcoin for increasing the fee paid in a transaction if the transaction becomes stuck in the memory-pool due to having a low fee relative to others in the pool; `replace-by-fee' (RBF) \cite{bip125} and `child-pays-for-parent' (CPFP). 

RBF is a signalling mechanism that alerts miners that a transaction may be replaced by an equivalent one with a higher fee. Miners opt-in to a policy of detecting these signals and updating their memory-pool if the replacement transaction is broadcast. They  have an incentive to claim the higher fee if it is there. Using this mechanism with an active covenant transaction is only possible if an appropriate alternative transaction with a higher fee has been prepared before the signing key is deleted. Therefore one strategy to dynamically allocate fees for funds controlled by covenant transactions is to prepare multiple versions of the same covenant transaction with a range of fees for different states of network congestion. For chains of covenant transactions, equivalent chains should be prepared which each signal with RBF. While RBF is particularly useful for already-broadcast unconfirmed transactions, a similar approach can be taken for un-broadcast transaction chains which don't rely on RBF signalling. That is to say, preparing in advanced alternative transaction chains with different fees would also enable dynamically allocating fees when it is time to broadcast (rather than \textit{after} broadcast), even without RBF signalling. However, for a chain of covenants of length $t$, if each covenant is prepared with $p$ variants, then $p^t$ covenant transactions must be prepared. This is because each covenant transaction is bound to a specific variant that it is spending from and so $p$ variants must be created for each new dependent covenant in the chain \textit{for all} $p$ variants of the previous covenant. This scales exponentially poorly in both storage and computation required for signature generation and verification.    


CPFP became the default transaction selection method in Bitcoin Core 0.13. Bitcoin's consensus rules enforce a strict ordering of transactions. However, transactions which depend on an unconfirmed transaction can be mined in the same block as its dependent ancestor (parent) so long as it appears later in the block. Thus, if an unconfirmed transaction has a small fee, but a child transaction has a fee large enough to cover itself and its unconfirmed parent transaction, miners have an incentive to include both into a block. If there is a chain of unconfirmed covenant transactions in the memory-pool, the final covenant transaction can be updated (before being broadcast) with a fee that is sufficiently large to pay for the whole chain. The final covenant transaction should thus be signed with the {\tt ANYONECANPAY|ALL} SIGHASH option in order to add a new input. If signed with {\tt ANYONECANPAY|SINGLE}, an additional change output may be added too. Recall however (see section \ref{subsec:ClassOfCovenants}) that signatures which don't commit to all inputs or all outputs can be replayed in alternative transactions, and observe that this enables transaction pinning attacks \cite{Pinning}. Here, an adversary would add a large change output to minimize the fee paid to miners, and would cause the entire unconfirmed transaction chain to be stuck, or `pinned', in the transaction processing queue. Another problem with relying on CPFP is that the whole chain of covenant transactions must be mined in the same block, which will not suit every application. Application designers should consider both CPFP and RBF, and the best approach might involve a combination of the two. The longer the chain of covenants, the more impractical it is to dynamically allocate fees. It is practical to dynamically allocate fees to very short covenant chains.

\subsection{Interactivity of a Multi-Party Deleted-key Covenant Protocol}

The requirement for short-lived ephemeral keys to enhance the security of the covenant mechanism necessitates an interactive protocol. To see this, consider the life-cycle of the ephemeral keys through the processes that must occur in a multi-party deleted-key covenant protocol: the deposit construction (where the ephemeral private keys are generated in order to derive the public keys for the covenant address), the distributed signing of the covenant transaction, and the deletion of the ephemeral private keys. Each of these parts requires communication and needs to happen quickly to reduce the life-cycle of the ephemeral keys to guarantee higher security for the covenant. That means that the participants involved must be online and ready, and makes enforcement process \textit{interactive}. If an application requires a tree of covenant transactions, then the set-up and deposit need not be repeated and additional communication rounds would not be required for the distributed signing, key deletion and storage of each additional covenant transaction since the operations could be batched. Despite these optimizations, the interactivity of the deleted-key covenant mechanism poses practical limitations to multi-party covenant protocols, especially as the number of participants increases, and most consequentially in adversarial contexts where any enforcer can stall the set-up process for all other enforcers. On the other hand, the interactivity may be an acceptable trade-off for the additional security that covenants can add to a given application.

\subsection{Multi-party ECDSA Threshold Signatures}


Currently, by default, bitcoin-core software only relays bare multi-signature transactions if $k \leq 3$ but may specify an upper limit of $k=20$ \cite{goldfeder2016threshold}. P2SH multi-signature scripts are bounded in size to limit denial-of-service attacks on validating nodes and the upper limit for such policies using this method is $k=15$. While on-chain threshold signature policies are significantly bounded, there are off-chain techniques that scale well enough to significantly surpass these bounds. For example, there has been a significant advance of proposals that enable distributed key generation and distributed signing in a multi-party setting for threshold signatures using ECDSA  \cite{Gennaro:2018:FMT:3243734.3243859,Lindell:2018:FSM:3243734.3243788,Canetti2020,Damgard2020,Gennaro2020, Gagol2020}. This is a promising area for enhancing the security, cost, scalability, and privacy properties of deleted-key covenants at the risk of introducing new cryptographic assumptions and reliance on additional software.

\section{Recovered-Key Covenants}
\label{sec:Recovered-Key Covenants}

Given some of the limitations and impracticalities of deleted-key covenants, it is worth considering how small modifications to the bitcoin protocol could enable alternative covenant mechanisms. This section introduces \textit{recovered-key covenants} which aren't currently possible, though with the soft-fork upgrade to the signature hash (SIGHASH) system proposed in BIP-118 they would be \cite{BIP118}. The following sections will introduce ECDSA public key recovery and outline the recovered-key covenant protocol with a security analysis. The factors that preclude recovered-key covenants are presented along with how they are solved by BIP-118. Finally, details of additional features that recovered-key covenants enable are given that are not possible with deleted-key covenants.   

\subsection{Elliptic Curve Public Key Recovery}

An interesting alternative to deleting private keys is to use elliptic curve (EC) public key recovery. When in possession of a signature and its message, it's possible to {\it compute} the public-key \cite{SEC1-ECC}. One chooses values for a {\it signature} and verifies that it is a valid EC point, instead of computing the signature according to the signing algorithm using the private key.  This is perhaps the best solution for key-deletion as it does not actually require deleting a key and mitigates concerns about whether or not the key was {\it really} deleted.

Details of the operation for ECDSA public key recovery are given in Section 4.1.6 of \cite{SEC1-ECC}. Let the curve's base point be $G$ and the group order be $n$. Let $\cdot$ represent the EC multiplication operation. Note that all finite field arithmetic is carried out modulo $n$. Given $e=L_b(H(m|r))$ as the lowest $b$ bits of the hashed message $m$, signature $(r,s)$ with $r$ as the $x$-component of an EC point $R$, and private key $p$, one can recover a public key $P=p\cdot G$ as
\begin{equation}
\label{eq:ECrecovery}
    P = r^{-1}(s\cdot R - e\cdot G)
\end{equation}
Several candidate public keys can be recovered from a given signature when mapping the chosen $(r,s)$ values to $P$\footnote{Given that the $x$-component is fixed, there are two choices (for the $y$-coordinate). Generally this would be multiplied by the cofactor of the curve (which is 1 for secp256k1).}. The signature should also be encoded in the DER format \cite{BIP62} and should be used to validate which candidate for $P$ is correct, using the signature verification algorithm.

\subsection{Protocol Specification}
\label{subsec:RecoveredKeyProtocolOutline}

Conceptually, this protocol is similar to deleted-key covenants. The same class of covenants are possible (section \ref{subsec:ClassOfCovenants}), joint commitments can be composed in a similar way (section \ref{subsec:Composability}), custodial power has the same determinants and the same process is required for proof-of-reserves (section \ref{sec:ProofOfReserves}). 
Unlike the deleted-key covenant protocol, a recovered-key covenant protocol requires no enforcer that is distinct from the depositor. The depositor, who is funding the covenant, can enforce the covenant and prove that it was enforced non-interactively. It is assumed that the set of custodial keys Q and the covenant transaction template are agreed upon before enacting the covenant enforcement. The commitment signature $(r,s)$ is valid over a message $m$ which is an appropriate digest of the covenant transaction (depending on which SIGHASH type is used). The address $A=A(P,Q)$ commits to the enforcement public key $P$ and to $Q$. The enforcement conditions for recovered-key covenants are $(a)$ commit to the transaction template with the commitment signature, and $(b)$ confirm the deposit transaction which funds an address $A=A(P,Q)$. When these conditions are met, the covenant transaction becomes \textit{active}.

\begin{enumerate}
    \item The depositor generates a `nothing up my sleeve' (NUMS) commitment signature $(r,s) \xleftarrow[]{} (1,1)$.
    \item The depositor verifies that $(r,s)$ is a valid EC point. If not, start again from step 1 and increase $r$ by 1.  
    \item The depositor computes a public key $P$ from the message $m$ and the commitment signature according to equation \ref{eq:ECrecovery} and iterates through candidates until one is verified by the signature verification algorithm.  
    \item The depositor constructs and broadcasts a deposit transaction that locks coins to an address $A = A(P,Q)$.
    \item The depositor sends the covenant transaction and its commitment signature to the set of custodians.
\end{enumerate}

The reason for step 2 is that mapping numbers sampled from a range to an elliptic curve often fails (roughly half of the time). This is a consequence of Hasse's theorem on elliptic curves which determines bounds for the number of points on a curve for a given finite field \cite{HasseTheorem}.  


\subsection{Security Analysis}

The security for the enforcement of a recovered-key covenant depends on how difficult it is to break its enforcement conditions. 
The commitment signature is secure since with ECDSA signatures the private key cannot be derived from knowledge of the signature or public key alone. This assumes that the EC discrete logarithm problem (ECDLP) is hard. The deposit confirmation is secure by assumption of the security of bitcoin consensus. Critically, there is no key deletion process and so the enforcement of recovered-key covenants is stricter than deleted-key covenants. Note also that there is no requirement for a secure RNG for the covenant enforcement (only for generation of custodial key pairs).

A critical advantage with recovered-key covenants is that it is possible to non-interactively prove that the private key is unknown. This means that the covenant enforcement requires no interaction and thus there are no multi-party risks involved. This covenant mechanism is practical in adversarial contexts where there is no trust between custodians and external parties who require proof-of-covenants. There is an implicit assumption that the depositor will honestly forward the necessary information to the set of custodians that are being paid. 
 
\vspace{0.1cm}
\noindent \textbf{Proposition:} By using NUMS quantities for the signature, one proves that it was generated without access to the private key. 

\vspace{0.1cm}
\noindent \textbf{Proof:} Let $P$, $(r,s)$ and $e$ be known quantities. The $s$-component of the signature is given by
\begin{equation}
    s = k^{-1}(e + r \cdot p)
\end{equation}
and rearranging for $q$ gives
\begin{equation}
\label{eq:privkey}
    p = r^{-1}(k \cdot s - e) 
\end{equation}
In addition to equation \ref{eq:privkey} the definition of $R$ is known,
\begin{equation}
    R = (r, R.y) = k \cdot G 
\end{equation}
but for any value of R, computing $k$ means solving the ECDLP which is computationally infeasible. Thus, equation \ref{eq:privkey} has two unknowns, $q$ and $k$, and the private key cannot be computed. The probability that a NUMS quantity signature was generated using a private key is negligible. Further assurance can be gained by having a challenger choose the NUMS quantities.  \vspace{-0.5cm}\begin{flushright} $\blacksquare$ \end{flushright} 

Consider now the risk that arises if many applications rely on the same NUMS quantity for $r$ in their covenant implementations. This amounts to a large scale re-use of a nonce $k$. Each recovered public key $P$ (and thus associated private key $p$) would be different, since it is determined by $e$ which is a function of the data in the covenant transaction. This leads to a system of equations of the same form as equation \ref{eq:privkey} where the $r$ and $s$ are known. While it is computationally infeasible to compute, if $k$ somehow did become known, the enforcement of all un-confirmed covenants could be broken. The message $e$ would be revealed as the covenant transaction is broadcast, and a swift attacker could then compute $p$ and broadcast an alternative theft transaction that replaces the covenant transaction and breaks its enforcement. For this reason it is recommended to use small and efficient NUMS quantities but not the same quantities in every case.

\subsection{Problem and Solution}

Unfortunately, EC key recovery covenant protocols are not possible on bitcoin today. The reason stems from two commitments made in the message (the covenant transaction) to the public key that funds are locked to (in the deposit transaction). It isn't possible to compute the public key without the message, but with the available SIGHASH types the message being signed requires a commitment to the public key.

First, with all current SIGHASH type options, the signature commits to an input which points to the TXID of the output that is being spent from (the TXID of the previous transaction). The TXID commits to that output which specifies the receiving address $A=A(P,Q)$, which commits to the enforcement public key through a hash function. The circular dependence on the public key cannot be resolved without breaking the hash function. Second, the signature also directly commits to the {\tt scriptPubKey} (or {\tt scriptCode} for Segregated Witness transactions) of the previous output, which is a function of the public key.

These circular dependencies would also be present if the Schnorr signature scheme was used in bitcoin. The current proposal for Schnorr signatures actually creates an additional commitment to the public key $P$ in the signature scheme itself, and prevents computation of $P$  (assuming the hash function cannot be broken) \cite{BIP-Schnorr}. For the signature $(r,s)$, the $s$-component is given by,
\begin{equation}
\label{eq:schnorr}
    s \cdot G = R + {\rm hash}(r|P|m)\cdot P \qquad \implies \qquad P = {\rm hash}(r|P|m)^{-1}(s \cdot G - R)
\end{equation}

\noindent Observe that $P$ is committed to by its inclusion in the hash function. This is done because linear transformations on the public key are used to commit to the Taproot spending conditions as well as in public derivations of keys under BIP-32 \cite{BIP-Taproot, BIP32}. If the signature algorithm did not commit to the public key, then an attacker could claim any Taproot spending conditions, and a watch-only wallet could maul signatures for one BIP-32 key tree to be valid for another. Adding this additional commitment to $P$ is important but it disables key recovery.


There has been a stream of soft-fork proposals that intend to remove commitments to the specific previous output from the signature, such that the signature for an input can be bound to any output with a compatible script. The idea is to introduce a new SIGHASH option, NO\_INPUT (via a Segregated Witness upgrade) or ANYPREVOUT (via a Tapscript upgrade), which allow dynamic binding of transactions to different outputs \cite{BIP118, BIP-anyprevout}. While both of these proposals remove the circular dependence on the public key, only ECDSA-based signatures enable covenants through public key recovery (unless the Schnorr signature algorithms are modified).  

The first of these proposals was presented in the original Lightning paper and in general the use cases presented by these proposals have focused on payment channels and off-chain protocols \cite{Poon2016, decker2018}. For example, this would allow a Lightning channel to be reformed by changing the input to which it refers. In this way, the balance in a channel can be modified without an on-chain transaction. This paper introduces another powerful use case: bitcoin covenants with tight security assumptions, practical multi-party protocols and support for dynamic fee allocation (see section \ref{sec:RKDynamicFee}). Enabling recovered-key covenants is a strong argument in favour of upgrading via the Segregated Witness transaction versioning system, rather than with the Tapscript versioning system.

\subsection{Dynamic Fee Allocation}
\label{sec:RKDynamicFee}

Since recovered-key covenant transactions don't commit to a specific previous output, the malleability of transaction IDs doesn't affect the safety of a chain of covenants. This is a critical point in favour of recovered-key covenants because it enables dynamic fee allocation, in particular with long covenant chains where transactions need to be confirmed at different times and in separate blocks (rather than all in the same block as is the case when relying on CPFP). Moreover, since malleability is less of a concern, the other SIGHASH types become safer to use and so the full class of possible covenants is more accessible (see section \ref{subsec:ClassOfCovenants}). 

The method for dynamic fee allocation relies on generating a signature for the covenant transaction input that uses the {\tt NO\_INPUT|ANYONECANPAY} SIGHASH type. Note that {\tt NO\_INPUT} implicitly commits to all outputs, so {\tt ALL} is not needed, and to none of the inputs. The signature commits to the conditions of the covenant, but allows for an additional fee-paying input to be added to the transaction at a later time. This additional input will yield a different transaction ID for the covenant, but if there is a dependent covenant transaction that is also signed with NO\_INPUT, it is still valid so long as the locking and unlocking scripts are compatible. Thus a covenant transaction can optionally be broadcast with a fee-input that is determined dynamically, that is, according to the \textit{current} state of network congestion.

\subsection{Disjoint Commitments}

It is possible to conditionally commit to a set of spending conditions by either preparing alternative covenant transactions that spend from the same output, or by using conditional logic in the output of the covenant transaction. The latter method is the same as with deleted-key covenants. The former method is slightly different and slightly less efficient than with deleted-key covenants. Consider figure \ref{fig:disjoint-covenant}, where the ephemeral enforcement key used to generate commitment signatures can be used for both transactions. This allows the deposit transaction to commit to both covenant transactions with the same public key. In contrast, with recovered-key commitment signatures, the private key is unknown. This means that each alternative covenant transaction has its own public key, and thus the deposit transaction must commit to each of these possible public keys in its output. 

\subsection{Privacy and Efficiency}

The commitment signature for a recovered-key covenant is identifiable through the {\tt NO\_INPUT} SIGHASH type. This will be public information when the covenant transaction is broadcast to the network. Where there is overlap between different protocols that use {\tt NO\_INPUT}, there will be an anonymity set about what kind of protocol or application the transaction was a part of. However, in general, there would not be much privacy for users of recovered-key covenants. 

Using NUMS quantities for a signature, such as $(r,s)=(1,1)$, is totally privacy breaking since it is clearly distinguishable from other signatures. On the other hand, it ensures that the size of the covenant transaction is minimised. A $(1,1)$ DER encoded signature is 9 bytes \cite{BIP66}. To obtain maximal privacy one could instead generate the signature using a hash function and two input seeds as
\begin{equation}
\label{eq:SHAsigs}
(r,s) \xleftarrow{} (SHA256(seed_r), SHA256(seed_s))    
\end{equation}

The outputted signature would then be indistinguishable from other transactions' signatures apart from the SIGHASH type used. Using a deterministic random oracle enables a proof that the commitment signature was generated without a known private key. Assuming that SHA256 is resistant to pre-image attacks, it would be impossible to discover the appropriate input seeds for a signature generated from a private key. To prove that a commitment signature was generated without a known private key, one simply needs to share their $seed_r$ and $seed_s$ and any challenger can verify that these yield the commitment signature $(r,s)$ by equation \ref{eq:SHAsigs}. Such a signature would consume more storage, with a total of 70 to 72 bytes when in DER format.

\section{Script-Based Covenants}
\label{sec:ScriptCovenants}




An obvious choice for adding additional functionality to bitcoin transactions is to upgrade the Script language. Several proposals have been made to add new OP\_CODEs to Script which enable more practical covenant functionality. The earlier proposals haven't been accepted by the bitcoin community as soft-fork upgrades \cite{moeser2016bitcoin, BIP-PUSHTXDATA, Covenants2}. These proposals have significant downsides as discussed by J. Rubin \cite{CovenantAlternatives}. Rubin is the author of the most recent proposal which yields relatively developer-friendly covenant Scripts, is simple to understand (and therefore analyse) and has an open-source implementation available on github \cite{BIP119, CTVImplementation}. The basic idea is to use a new OP\_CODE called {\tt OP\_CHECKTEMPLATEVERIFY} (CTV) when generating an UTxO and commit to a transaction template that must be matched by any transaction that spends this output. Specifically, the commitment is a hash of the serialized version, locktime, scriptSigs hash (if there are any non-null scriptSigs), number of inputs, sequences hash, number of outputs, outputs hash, and currently executing input index. A simple example of a deposit transaction's CTV locking script where the \textit{commitment hash} is derived from a covenant transaction and the UTxO is guarded by a private key for an associated custodial public key $Q$ takes the form: 
\begin{verbatim}
            <Q> OP_CHECKSIGVERIFY <Commitment Hash> OP_CHECKTEMPLATEVERIFY
\end{verbatim} 

The class of possible covenants is determined by the commitment hash type. Initially there will be only one type, the {\tt StandardTemplateHash}, which defines what transaction data are input to the commitment hash. 
This template doesn't exactly map on to any of the SIGHASH types that determine what parts of a transaction the commitment signature(s) apply to with deleted-key and recovered-key covenants. Both the template hashes and the SIGHASH types that are available now could (in theory) be updated in future soft-forks to broaden the class of possible covenants. In contrast to commitment signatures (with any SIGHASH type), the {\tt StandardTemplateHash} always commits to the number of inputs (and outputs). This has consequences for the methodology of dynamic fee allocation with covenant chains (see section \ref{sec:SBDFA}). The main functional differences between recovered-key and script-based covenants are determined by the difference in precisely what the commitment signature and commitment hash can commit to. 

\subsection{Security Analysis}

The enforcement conditions for a script-based covenant are $(a)$ commit to a covenant transaction with the commitment hash, and $(b)$ confirm the deposit transaction. As with recovered-key covenants, script-based covenants derive many advantages from the fact that there is no dependence on a secure key deletion process. Instead, the security of script-based covenants assumes that bitcoin's consensus protocol operates correctly, and that the hash function used to generate the commitment hash is secure. 

Script-based covenant protocols have no required interaction for their enforcement, which is an advantage when compared with deleted-key covenants. Only the depositor participates in the enforcement (provided that a set of custodial keys and the covenant conditions have been agreed upon). Thus, script-based covenants are far more practical in adversarial multi-party contexts. As with recovered-key covenants, there is an implicit assumption that the depositor acts honestly and forwards the covenant transaction data to the custodians.

\subsection{Performance and Cost}

A key difference between script-based and recovered-key covenants comes from the fact that the former requires computing a hash output and the latter requires verifying a signature. Signature verification is significantly more costly to validating nodes in the network. Moreover, the size of (deposit and covenant) transactions affects the required transaction fees for covenant users and the blockchain storage cost. The size of the commitment hash is 32 bytes and a simple CTV script is 34 bytes. 

Note that an important difference between recovered-key and script-based covenants would emerge if the Schnorr signature scheme is accepted as a soft-fork upgrade in bitcoin \cite{BIP-Schnorr}. Script-based covenants could make use of signature and public key aggregation techniques to significantly improve the storage costs of the custodial access policy.

\subsection{Proof-of-Reserves and Proof-of-Covenants}

With Script-based covenants, the ability to prove that an output is bound by a covenant does not require the presence of the challenger during set-up, but can be done at any time after the transaction generating the covenant-bound output is confirmed on the blockchain. This makes auditing custody operations that rely on covenants much simpler for both the custodian and the auditor than with deleted-key covenants. One can prove that they control a private key for an associated public key by signing an invalid transaction, according to the proof-of-reserves protocol \cite{BIP127}. Proving that an output (which is publicly visible on the blockchain) is bound by a specific covenant amounts to revealing the transaction template that was committed to and demonstrating that the commitment hash derives from it. 

\subsection{Dynamic Fee Allocation}
\label{sec:SBDFA}

Script-based covenants allow for the ability to dynamically allocate fees to covenant transactions at the time of broadcast. As a demonstration of this, consider using CTV and assume the use of Segregated Witness addresses such that any signatures aren't included in the commitment hash (refer to the specification \cite{BIP119}). An output can commit to being spent by two variants of a covenant transaction, Tx and Tx$'$, where Tx$'$ has an additional input for paying fees. An example locking script (where the custodial keys are ommitted) takes the following form,
\begin{verbatim}
      OP_IF <Tx Commitment Hash> OP_ELSE <Tx’ Commitment Hash> OP_ENDIF OP_CTV
\end{verbatim}

The fee input may spend from any output, including outputs that didn't exist until after the transaction was committed to. This means that a chain of covenants that use CTV don't require non-malleable intermediate transaction IDs because the specific outputs to be spent (the objects that refer to previous transactions' IDs) aren't committed to. Thus, the appropriate variant of each intermediate planned transaction can be chosen at the time of broadcast (or even replaced after broadcast using RBF) without invalidating subsequent dependent covenants. Dynamic fee allocation simplifies the implementation of applications that require longer covenant chains, and will likely reduce the overall cost of transaction fees for such applications and alleviate the need to overpay fees to ensure that transactions are processed. 

Using the SIGHASH type {\tt ANYONECANPAY|ALL} is slightly restricted with CTV since the number of inputs are included in the commitment hash and thus the optionality of adding an input fee is lost. Instead, an additional input must be added to a covenant transaction that has an input signed with {\tt ANYONECANPAY|ALL}. If one relies on conditional commitment hashes in the script, this manifests as a larger byte size for long covenant chains where dynamic fee allocation is needed. However, pruning techniques could be implemented for non-essential script data. To compare with recovered-key covenants, the cost of verification might be a more important factor for the scalability of bitcoin than the storage costs.

\section{Conclusion}
\label{sec:Conclusion}

A bitcoin \textit{covenant} is a mechanism to enforce restrictive conditions on future bitcoin transactions. Covenants in general are a powerful tool with applications in custody and payment protocols. This paper introduced two covenant mechanisms; \textit{deleted-key} and \textit{recovered-key} covenants. The former is possible today, while the latter isn't and requires a soft-fork upgrade such as BIP-118 \cite{BIP118}. 

Deleted-key covenants are enforced by a key deletion process. A range of security models are possible for this process, such as relying on distributed signing devices with secure deletion functions or relying on disposable signing devices that are destroyed once the signature has been generated. This mechanism is subject to an inherent trade-off between security and convenience; the more secure the covenant enforcement (for example through the use of numerous off-line signing devices), the more inconvenient it is to activate the covenant. Since it is not possible for one party to prove to another that they have deleted a key, any party who requires a proof that the covenant is being enforced must participate in the enforcement process. This is a severe limitation for covenant-based applications that need to be audited or that operate in adversarial contexts. Deleted-key covenants are more likely to be practical in self-custody and joint-custody protocols than payment protocols due to risks involved in adversarial multi-party settings that arise from an interactive enforcement process. This paper formalized the security properties of deleted-key covenants and discussed critical factors for the design of covenant-based applications including; the class of possible covenants, countermeasures for the risks involved when composing multiple covenant and deposit transactions, and techniques for dynamic fee allocation with short covenant chains (long chains are impractical). In particular, deleted-key covenants are a promising tool for custody protocols since they can restrict the ways in which coins can be spent, and thus can be used to create choke-points, rate-limits, and sequential barriers that an adversary must overcome in order to steal funds. 

Recovered-key covenants have a non-interactive enforcement process and a non-interactive method for proving that funds are bound by a covenant. The security of their enforcement requires no assumption about a key deletion process nor of the secure generation of a private key. Thus, they are more practical than deleted-key covenants in all contexts (including self-custody, joint-custody and payment protocols). Enabling recovered-key covenants requires a soft-fork upgrade to create a new signature hash (SIGHASH) type that allows dynamic binding of commitment signatures to different previous outputs. Such an upgrade has been proposed elsewhere as it simplifies off-chain protocols \cite{Poon2016, BIP118, decker2018, BIP-anyprevout}. With recovered-key covenants, dynamic fee allocation is superior since it is safe to maul an intermediate transaction in a covenant chain. This means additional fees can be added to any transaction in the chain without rendering any dependent transactions invalid.  


This work gave a comparative analysis of deleted-key, recovered-key and Script-based covenants (proposed elsewhere). The main results are summarised in table \ref{tab:table}. Recovered-key and script-based covenants are mostly functionally equivalent and so the advantages that recovered-key covenants have over deleted-key covenants also applies to Script-based covenants. If either were enabled by their required soft-fork upgrade then a new domain of practical covenant-based protocols could emerge. Understanding precisely what utility is gained from such upgrades is key to their progress.

The main barrier to recovered-key and Script-based covenants is the difficulty of reaching consensus for a soft-fork of Bitcoin. Bitcoin is a complex adaptive system\footnote{Bitcoin is an open network of independent decision-making agents who self-organize by participating according to a protocol (which is instanciated in code). The protocol is dynamic and changes with the actions of its constituent agents who thus affect the emergent functionality of bitcoin.} with many interacting parts and there are systemic risks with every modification of bitcoin's code-base and protocol. It is difficult to analyse those risks and it would be hubris to claim that there are no unknown risks being introduced. Moreover, consensus must be reached among developers \textit{and} non-technical users of Bitcoin. This results in an increasingly difficult social co-ordination problem as more users buy into the system. In practice there are far more proposals to modify Bitcoin than reviewers have capacity for and most aren't accepted \cite{BIPS}.



\section*{Acknowledgements}

We thank Professor McBurney (King's College London), Sam Abbassi (Fidelity Center for Applied Technology), Jeremy Rubin and Madars Visra (MIT Media Lab) for constructive feedback on this work.

\section*{Funding}

Funding is gratefully acknowledged under a UK EPSRC-funded GTA Award through King's College London, and from EPSRC Research Grant EP/P031811/1, the Voting Over Ledger Technologies (VOLT) Project. This work was also supported by Fidelity Center for Applied Technology. 

\bibliography{sample}

\end{document}